\documentclass[twocolumn,aps,nofootinbib]{revtex4-1}

\usepackage{graphicx}
\usepackage{epstopdf}
\usepackage{latexsym}
\usepackage{amsmath}
\usepackage{amssymb}
\usepackage{color}
\usepackage{mathrsfs}

\usepackage[center]{subfigure}

\begin{document}

 \newcommand{\bq}{\begin{equation}}
 \newcommand{\eq}{\end{equation}}
 \newcommand{\bqn}{\begin{eqnarray}}
 \newcommand{\eqn}{\end{eqnarray}}
 \newcommand{\nb}{\nonumber}
 \newcommand{\lb}{\label}
 \newcommand{\be}{\begin{equation}}
\newcommand{\en}{\end{equation}}
\newcommand{\PRL}{Phys. Rev. Lett.}
\newcommand{\PL}{Phys. Lett.}
\newcommand{\PR}{Phys. Rev.}
\newcommand{\CQG}{Class. Quantum Grav.}

\title{Quasinormal modes and echo effect of cylindrical anti-de Sitter black hole spacetime with a thin shell}

\author{Kai Lin}\email{lk314159@hotmail.com}

\affiliation{Hubei Subsurface Multi-scale Imaging Key Laboratory, School of Geophysics and Geomatics, China University of Geosciences, Wuhan 430074, Hubei, China}

\date{\today}

\begin{abstract}
This paper investigates the quasinormal mode (QNM) vibrations of a rotating cylindrical black hole (or black string) spacetime that is surrounded by a thin shell rotating synchronously with the black string's axis. The existence of the thin shell leads to a piecewise metric of the black hole spacetime beyond the horizon, which is divided into two stationary spacetime parts by the radius of the thin shell. As a result, the potential function $V(r)$ of the QNM equation is also discontinuous. To solve the QNM equation with the discontinuous potential function, we propose two methods, matrix method and generalized Horowitz-Hubeny Method. We find that the influence of the thin shell can reduce the QNM frequency of the black string while alleviating their amplitude decay rate. Our suggested method can be easily applied to other QNM calculations of black hole spacetime with discontinuous potential function, thus facilitating investigations into more intricate and realistic black hole spacetimes, such as those with accretion disks. Additionally, the finite difference method is employed to investigate the spacetime too. This analysis discloses a substantial gap in the potential function when the thin shell's mass and charge achieve sufficiently high values, resulting in the outer spacetime nearing gravitational collapse and extreme black hole scenarios. Within this gap, the QNM wave displays oscillations, producing an echo effect. Moreover, it is established that the closeness of the spacetime to the collapse threshold and charge extremality have positive correlation with the beat interval of this echo.
\end{abstract}


\maketitle

\section{Introduction}
\renewcommand{\theequation}{1.\arabic{equation}} \setcounter{equation}{0}

In 2015, a momentous discovery by The Laser Interferometer Gravitational-Wave Observatory (LIGO) verified the existence of these elusive entities by detecting gravitational waves emanating from a binary black hole coalescence\cite{LIGO1,LIGO2,LIGO3,LIGO4,LIGO5,LIGO6,LIGO7,LIGO8,LIGO9}. The signal from such a cataclysmic event comprises three phases: first, an inspiral phase when the two black holes gravitate towards each other in a tightening orbit; then, a merger phase when they collide and coalesce into one massive body with a radiant outburst; finally, a ringdown phase when the remaining perturbations dissipate as the final black hole reaches equilibrium. For a stable black hole spacetime, ringdown reflects the dynamical evolution of the black hole under a minor disturbance. A potent technique for probing this phase is to apply perturbation theory to derive quasinormal mode (QNM) that depict how oscillations decay around a black hole. More broadly, any quantum field theory that adheres to Lorentz symmetry can describe the evolution of quantum field fluctuations in a curved black hole spacetime background.

Several numerical techniques have been developed to calculate the QNM frequency, such as WKB approximation\cite{WKB1,WKB2,WKB3,WKB4}, continued fraction method (CFM)\cite{CFM}, asymptotic iteration method (AIM)\cite{AIM1,AIM2,AIM3}, Horowitz-Hubeny method (HHM)\cite{HHM}, matrix method (MM)\cite{Lin1,Lin2,Lin3,Lin4,Lin5,Lin6,Lin7}, finite difference method (FDM)\cite{FDM1,FDM2,FDM3,FDM4,FDM5} and so on. These methods can describe the dynamical evolution of various black holes, such as static, rotating, anti-de Sitter(AdS) or de Sitter cases. However, many methods (except MM and FDM) can only solve the decoupled QNM equation, so we must use various tricks to simplify them. This implies that the higher spin case is more complex for simplifying the QNM master equation.

Black holes are theoretically simple but practically intricate entities in the real universe. Their formidable gravitational field draws and amasses various forms of matter beyond their event horizon. Hence, to capture the actual spacetime geometry of a black hole, one must incorporate external factors such as accretion disks and dark matter. As a result, real black holes spacetime may not conform to a smooth mathematical representation. A spacetime metric that entails a discontinuous function would also imply a non-smooth potential function for the QNM equation. It is a novel challenge to explore QNM of realistic black hole spacetime.

This paper explores a simple example of a nonsmooth spacetime metric: an outer thin shell with mass and charge enclosing a rotating anti-de Sitter (AdS) black string. The thin shell shares its axis of symmetry with the event horizon of the black hole and spins at an equal angular velocity as the black string. We will use three numerical methods to investigate the QNM dynamics in this non-smooth spacetime. In Sec. II, we present and simplify the metric for this thin-shell black string configuration, and then obtain the QNM equation for a scalar field in this background geometry. In Sec. III, we describe three numerical methods for solving this equation: Matrix Method, Generalized Horowitz-Hubeny Method, and Finite Difference Method. We apply these methods for computation in Sec. IV. We also find that: when the mass of the thin shell approaches critical mass of gravitational collapse and its charge approaches extremality, there is a large gap in the potential function that causes an echo effect. We analyze this echo effect further in Sec. IV. Sec. V contains some discussions and conclusions.

\section{Quasinormal Modes Master Equation of Rotating Black String with Thin Shell}
\renewcommand{\theequation}{2.\arabic{equation}} \setcounter{equation}{0}

According to \cite{metric}, the rotating charged cylindrical anti-de Sitter black hole with $\alpha^2\equiv-\Lambda/3$ has the following form:

\bqn
\lb{Metric1}
ds^2&=&-f(r)\left(\Xi d\bar{t}-\frac{\bar{\omega}}{\alpha^2}d\bar{\varphi}\right)^2+\frac{dr^2}{f(r)}\nb\\
&&+r^2\left(\Xi d\bar{\varphi}-\bar{\omega} d\bar{t}\right)^2+\alpha^2 r^2 d\bar{z}^2,\nb\\
f(r)&=&\alpha^2 r^2-\frac{4M}{\alpha r}\left(1-\frac{3}{2}\alpha^2 a^2\right)+\frac{4Q^2}{\alpha^2 r^2}\frac{2-3\alpha^2 a^2}{2-\alpha^2 a^2}\nb\\
&\equiv&\alpha^2 r^2-\frac{4M_\text{eff}}{\alpha r}+\frac{4Q^2_\text{eff}}{\alpha^2 r^2},
\eqn
where $M$, $Q$ and $a$ represent the mass per unit length, the linear charge density and the angular momentum per unit mass of the black string respectively. $M_\text{eff}$, $Q_\text{eff}$, $\Xi$ and $\bar{\omega}$ are defined by
\bqn
\lb{Metric2}
M_\text{eff}&\equiv&M\left(1-\frac{3}{2}\alpha^2 a^2\right),\nb\\
Q^2_\text{eff}&\equiv&Q^2\frac{2-3\alpha^2 a^2}{2-\alpha^2 a^2},\nb\\
\Xi&\equiv&\sqrt{\frac{2-\alpha^2 a^2}{2-3\alpha^2 a^2}},\nb\\
\bar{\omega}&\equiv&\frac{\sqrt{2}\alpha^2 a}{\sqrt{2-3\alpha^2 a^2}},
\eqn
and the cylindrical black hole horizon $r_h(>0)$ that satisfies $f(r_h)=0$. It is obvious that, at the event horizon, the angular velocity of black string is given by $\Omega_H\equiv\left.d\bar{\varphi}/d\bar{t}\right|_{r=r_h}=\bar{\omega}/\Xi$.

In this paper, we consider a rotating black string with a cylindrical thin shell at $r_s(>r_h)$ that rotates around the black string with the same angular momentum per unit mass $a$. This allows us to describe the effective mass and effective charge of the black string spacetime with thin shell as
\bqn
\lb{Metric3}
M_\text{eff}&=&
\left\{
\begin{array}
{cc}
M_\text{in} & r_s\ge r\ge r_h \\
M_\text{out}(\equiv M_\text{in}+M_\text{shell}\ge M_\text{in}) & r\ge r_s
\end{array}
\right.\nb\\
Q_\text{eff}&=&
\left\{
\begin{array}
{cc}
Q_\text{in} & r_s\ge r\ge r_h \\
Q_\text{out}(\equiv Q_\text{in}+Q_\text{shell}) & r\ge r_s
\end{array}
\right.
\eqn
Upon introducing the coordinate transformation
\bqn
\lb{Metric4}
t&=&\gamma \bar{t}-\frac{\bar{\omega}}{\alpha^2}\bar{\varphi}\nb\\
\varphi&=&\gamma \bar{\varphi}-\bar{\omega} \bar{t}\nb\\
z&=&\alpha\bar{z},
\eqn
the metric is simplified as
\bqn
\lb{Metric5}
ds^2=-f(r)dt^2+\frac{dr^2}{f(r)}+r^2d\varphi^2+r^2 dz^2,
\eqn
which yields an effective staticlike black string metric. We illustrate this in Fig.1.
\begin{figure}[tbp]
\centering
\includegraphics[width=0.7\columnwidth]{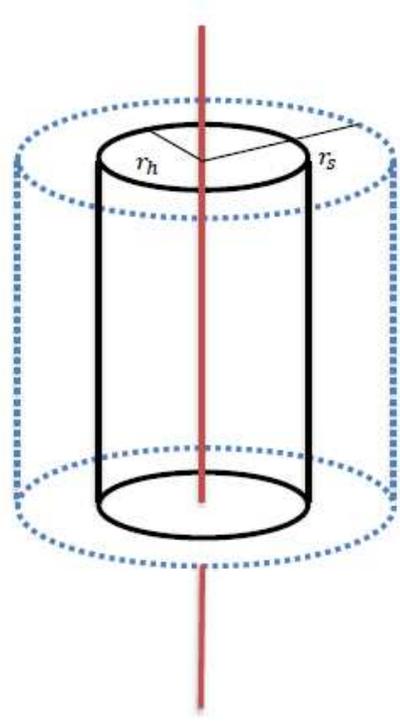}
\caption{The effective static-like black string with a thin shell. The solid line indicates the horizon ($r_h$) of the black hole, and the dotted line marks the location ($r_s$) of the thin shell with mass $M_\text{shell}$ and electrical charge $Q_\text{shell}$. From the figure, we observe that $M_\text{eff}=M_\text{in}$ and $Q_\text{eff}=Q_\text{in}$ hold in region $r_s\ge r\ge r_h$, while $M_\text{eff}=M_\text{out}$ and $Q_\text{eff}=Q_\text{out}$ hold in region $r\ge r_s$.}
\lb{Fig1}
\end{figure}

Without loss of generality, we set $\alpha=1$ and rewrite $f(r)$ as
\bqn
\lb{Metric6}
f(r)&=&
\left\{
\begin{array}
{cc}
f_\text{in}(r) & r_s\ge r\ge r_h \\
f_\text{out}(r)& r\ge r_s
\end{array}
\right.\nb\\
f_\text{in}(r)&\equiv&r^2-\frac{4M_\text{in}}{r}+\frac{4Q^2_\text{in}}{r^2}\nb\\
&=&\left(1-\frac{r_h}{r}\right)\left(1-e_\text{in}\frac{r_h}{r}\right)\left[\left(1+e_\text{in}\right)r_h r\right.\nb\\
&&\left.+r^2+\left(1+e_\text{in}+e_\text{in}^2\right)r_h^2\right],\nb\\
f_\text{out}(r)&\equiv&r^2-\frac{4M_\text{out}}{r}+\frac{4Q^2_\text{out}}{r^2}\nb\\
&=&\left(1-\frac{r_p}{r}\right)\left(1-e_\text{out}\frac{r_p}{r}\right)\left[\left(1+e_\text{out}\right)r_p r\right.\nb\\
&&\left.+r^2+\left(1+e_\text{out}+e_\text{out}^2\right)r_p^2\right],
\eqn
where $r_p=c_\text{out} r_h$ represents the gravitational radius of the AdS spacetime with a thin shell, and it means matter distributed within this radius will inevitably be attracted into the black hole, so the position of the thin shell must satisfy the condition $\frac{r_s}{r_h}(\equiv b)> c_\text{out} >1$. In addition, the spacetime may be charged, but the charge of black hole spacetime has an upper limit because of the cosmic censorship principle. We set the parameters $e_\text{in}(\in [0,1))$ and $e_\text{out}(\in [0,1))$ depending on the charge and mass of the black hole spacetime with a thin shell. Hence, the relation of $M_\text{in}$, $Q_\text{in}$, $M_\text{out}$, $Q_\text{out}$ and $r_h$, $r_p$, $e_\text{in}$, $e_\text{out}$ follows from
\bqn
\lb{Metric7}
M_\text{in}&=&\frac{1}{4}\left(1+e_\text{in}+e_\text{in}^2+e_\text{in}^3\right)r_h^3,\nb\\
M_\text{out}&=&\frac{1}{4}\left(1+e_\text{out}+e_\text{out}^2+e_\text{out}^3\right)r_p^3,\nb\\
&=&\frac{1}{4}\left(1+e_\text{out}+e_\text{out}^2+e_\text{out}^3\right)c_\text{out}^3r_h^3,\nb\\
Q^2_\text{in}&=&\frac{1}{4}\left(e_\text{in}+e_\text{in}^2+e_\text{in}^3\right)r_h^4,\nb\\
Q^2_\text{out}&=&\frac{1}{4}\left(e_\text{out}+e_\text{out}^2+e_\text{out}^3\right)r_p^4,\nb\\
&=&\frac{1}{4}\left(e_\text{out}+e_\text{out}^2+e_\text{out}^3\right)c_\text{out}^4r_h^4.
\eqn

Assuming a massless scalar perturbation, we obtain the minimally coupled scalar wave equation as $g^{\mu\nu}\Phi_{;\mu\nu}=0$. By setting $\Phi=\frac{1}{r}\Psi(t,r)\exp\left(ikz+iL\varphi\right)$, we can derive the radial master equation for scalar quasinormal modes as
\bqn
\lb{QNMsA}
f(r)\frac{\partial}{\partial r}\left(f(r)\frac{\partial \Psi}{\partial r}\right)-\frac{\partial^2 \Psi}{\partial t^2}-V(r)\Psi=0,
\eqn
where
\bqn
\lb{QNMsB}
V(r)&=&f(r)U(r),\nb\\
U(r)&=&\frac{L^2+k^2}{r^2}+\frac{f'(r)}{r}.
\eqn
Then, by using the tortoise coordinate $r_*=\int\frac{dr}{f(r)}$, we set $\Psi(t,r)=R(r)\exp\left[-i\omega (t+r_*)\right]$, so Eq.(\ref{QNMsA}) becomes
\bqn
\lb{QNMsC}
f(r)R''(r)+\left[f'(r)-2i\omega\right]R'(r)-U(r)R(r)=0.\nb\\
\eqn
Since $M_\text{eff}$ and $Q_\text{eff}$ vary inside and outside the thin shell, we reformulate the preceding equations as
\bqn
\lb{QNMsD}
&&\text{for } r_s\ge r \ge r_h\nb\\
&&f_\text{in}(r)R''_\text{in}(r)+\left[f'_\text{in}(r)-2i\omega\right]R'_\text{in}(r)\nb\\
&&~~~~~-U_\text{in}(r)R_\text{in}(r)=0,\nb\\
&&\text{for } r \ge r_s\equiv br_h\nb\\
&&f_\text{out}(r)R''_\text{out}(r)+\left[f'_\text{out}(r)-2i\omega\right]R'_\text{out}(r)\nb\\
&&~~~~~-U_\text{out}(r)R_\text{out}(r)=0.
\eqn

The boundary condition and the join condition requires
\bqn
\lb{QNMsE}
R_\text{in}(r_h)=\text{Constant}\nb\\
R_\text{out}(r\rightarrow \infty) \rightarrow  0\nb\\
\frac{R'_\text{in}(r_s)}{R_\text{in}(r_s)}=\frac{R'_\text{out}(r_s)}{R_\text{out}(r_s)}.
\eqn

To simplify our analysis, we consider the scenario where $r_h=1$ and $e_\text{in}=0$ hold. However, the QNM equation is difficult to solve analytically due to the complex potential function $V(r)$, so we propose three numerical methods to tackle the problem with a non-continuous potential function.

\section{Numerical Method}
\renewcommand{\theequation}{3.\arabic{equation}} \setcounter{equation}{0}

Typically, the conventional numerical methods are inadequate to handle the QNM equation with a non-continuous potential function, so we need to enhance the standard method. In this section, we present three numerical methods that can address the preceding problem, namely Matrix Method (MM), Generalized Horowitz-Hubeny Method (GHHM) and Finite Difference Method (FDM).

\subsection{Matrix Method}
To transform the coordinates in region $r_s\ge r\ge r_h$, we apply a mapping function
\bqn
\lb{MM1}
x&=&\frac{r-r_h}{r_s-r_h},\nb\\
w&=&\frac{\omega}{r_h}
\eqn
so that $x(r_h)=0$ and $x(r_s)=1$, and the QNM equation becomes
\bqn
\lb{MM2}
&&B_2(x)R''_\text{in}(x)+B_1(x)R'_\text{in}(x)+B_0(x)R_\text{in}(x)=0\nb\\
&&B_2(x)=x\left(1-x+bx\right)\left[3+3(b-1)x+(b-1)^2x^2\right],\nb\\
&&B_1(x)=3+9(b-1)x+12(b-1)^2x^2+8(b-1)^3x^3\nb\\
&&~~~~~~+2(b-1)^4x^4-2iw\left(1-x+bx\right)^3,\nb\\
&&B_0(x)=(b-1)\left[(1-x+bx)(L^2+k^2)r_h^{-2}+3\right.\nb\\
&&~~~~~~\left.+6(b-1)x+6(b-1)^2x^2+2(b-1)^3x^3\right].
\eqn

To transform the coordinates in region $r\ge r_s$, we apply a mapping function
\bqn
\lb{MM3}
y&=&1-\frac{r_s}{r},\nb\\
w&=&\frac{\omega}{r_h}
\eqn
so that $y(r_s)=0$ and $y(r\rightarrow\infty)=1$, and the QNM equation is
\bqn
\lb{MM4}
&&C_2(y)R''_\text{out}(y)+C_1(y)R'_\text{out}(y)+C_0(y)R_\text{out}(y)=0\nb\\
&&C_2(y)=\left(b-c_\text{out}+c_\text{out}y\right)(y-1)^2\left[b^3\right.\nb\\
&&~~~~~~+c_\text{out}^3e_\text{out}(1+e_\text{out}+e_\text{out}^2)(y-1)^3\nb\\
&&~~~~~~\left.+bc_\text{out}^2(y-1)^2+b^2c_\text{out}(1-y)\right]\nb\\
&&C_1(y)=c_\text{out}^3(y-1)^4\left[3b(1+e_\text{out}+e_\text{out}^2+e_\text{out}^3)\right.\nb\\
&&~~~~~~\left.+4c_\text{out}e_\text{out}(1+e_\text{out}+e_\text{out}^2)(y-1)\right]\nb\\
&&~~~~~~-2ib^3(y-1)^2w\nb\\
&&C_0(y)=b^2(L^2+k^2)r_h^{-2}(y-1)^2-2b^4\nb\\
&&~~~~~~+bc_\text{out}^3(1+e_\text{out}+e_\text{out}^2+e_\text{out}^3)(y-1)^3\nb\\
&&~~~~~~+2c_\text{out}^4e_\text{out}(1+e_\text{out}+e_\text{out}^2)(y-1)^4
\eqn

The join condition becomes
\bqn
\lb{MM5}
\frac{R_\text{in}'(x=1)}{(b-1)R_\text{in}(x=1)}=\frac{R_\text{out}'(y=0)}{bR_\text{out}(y=0)}=\lambda,
\eqn
or
\bqn
\lb{MM6}
R_\text{in}'(x=1)-\lambda (b-1)R_\text{in}(x=1)&=&0,\nb\\
R_\text{out}'(y=0)-\lambda b R_\text{out}(y=0)&=&0
\eqn

Based on the Matrix Method, we first divide the region $x\in[0,1]$ (or $y\in[0,1]$) into a set of discrete points $\{x_1=0,x_2,\cdot\cdot\cdot,x_n=1\}$ and obtain a corresponding set of functions $\chi=[R(x_0),R(x_1),\cdot\cdot\cdot,R(x_n)]^T$. Then we apply a high-accuracy difference method to discretize the preceding differential equation. Such methods include the Runge-Kutta method, the differential quadrature method and a non-grid-based interpolation scheme in \cite{Lin1}. In this paper, we adopt the latter scheme for computation and express the derivative of order $m$ for $R^{(m)}(x_i)$ at point $x_i$ as a linear combination of all values of $R(x_j)$ in $\chi$. Hence,we substitute $R^{(m)}(x_i)$ with $\tilde{M}^{\{m\}}_{i}\chi_i$, resulting in
\bqn
\lb{MM7}
&&\text{for }r_s\ge r\ge r_h~~~{\mathbb M}_\text{in}\chi_\text{in}\equiv\nb\\
&&~~~~~\left(B_2\tilde{M}^{\{2\}}+B_1\tilde{M}^{\{1\}}+B_0\tilde{M}^{\{0\}}\right)\chi_\text{in}=0,\nb\\
&&\text{for }r\ge r_s~~~{\mathbb M}_\text{out}\chi_\text{out}\equiv\nb\\
&&~~~~~\left(C_2\tilde{M}^{\{2\}}+C_1\tilde{M}^{\{1\}}+C_0\tilde{M}^{\{0\}}\right)\chi_\text{out}=0,
\eqn
where $\tilde{M}_i$ denotes the $i$-th row of a square matrix with details given in \cite{Lin1}. The boundary condition stipulates that $R_\text{in}(x=0)$  remains constant and $R_\text{out}(y=1)$ disappears. The join condition needs to be discretized as
\bqn
\lb{MM8}
\tilde{M}^{\{1\}}_n-\lambda (b-1)\tilde{M}^{\{0\}}_n&=&0,
\eqn
and
\bqn
\lb{MM9}
\tilde{M}^{\{1\}}_1-\lambda b \tilde{M}^{\{0\}}_1&=&0.
\eqn
So we substitute Eq.(\ref{MM8}) and Eq.(\ref{MM9}) for the final row of ${\mathbb M}_\text{in}$ and the initial row of ${\mathbb M}_\text{out}$ respectively, and then modify the final row of ${\mathbb M}_\text{out}$ as $(0,0,\cdot\cdot\cdot,1)$ to prevent potential numerical singularity. As a result, the matrix ${\mathbb M}_\text{in}$ and ${\mathbb M}_\text{out}$ transform into $\tilde{\mathbb M}_\text{in}$ and $\tilde{\mathbb M}_\text{out}$ respectively. Finally, to solve for the preceding eigenmatrix, we equate the determinant of the matrix with zero:
\bqn
\lb{MM10}
\det{\tilde{\mathbb M}_\text{in}(w,\lambda)}&=&0,\nb\\
\det{\tilde{\mathbb M}_\text{out}(w,\lambda)}&=&0.
\eqn
The eigenvalue $\omega$ and $\lambda$ could be obtained by the command ``FindRoot" or ``NSolve" in Mathematica software.

\subsection{Generalized Horowitz-Hubeny Method}

In \cite{HHM}, Horowitz and Hubeny introduced a novel approach to compute the QNM frequency in AdS spacetime. The approach resembles the Continued fraction method, but it has an edge over it as it does not derive the recurrence relation for the coefficient function. Thus, it does not demand that the coefficient function of the QNM equation be a rational expression. However, when dealing with a metric with a piecewise function, the initial HH method falls short and needs to be generalized here.

By using the coordinate transformation $z=1-\frac{r_h}{r}$ and $w=\frac{\omega}{r_h}$, Eq.(\ref{QNMsC}) becomes
\bqn
\lb{HH1}
&&A_2(z)R''(z)+A_1(z)R''(z)+A_0(z)R(z)=0\nb\\
A_2(z)&=&(1-c+cz)(z-1)^2\left[1+c+c^2(z-1)^2\right.\nb\\
&&\left.+c^3e(1+e+e^2)(z-1)^3-cz\right],\nb\\
A_1(z)&=&c^3(z-1)^4\left[3+(3+4cz-4c)(e+e^2+e^3)\right]\nb\\
&&-2iw(z-1)^2,\nb\\
A_0(z)&=&2c^4(e+e^2+e^3)(z-1)^4\nb\\
&&+c^3(1+e+e^2+e^3)(z-1)^3-2\nb\\
&&-(L^2+k^2)r_h^{-1}(z-1)^2.
\eqn
According to the boundary condition, we set
\bqn
\lb{HH2}
\frac{A_1(z)}{A_2(z)}&=&\sum^\infty_{k=-1}\hat{B}_kz^k\nb\\
\frac{A_0(z)}{A_2(z)}&=&\sum^\infty_{k=-1}\hat{C}_kz^k\nb\\
R_\text{in}(z)&=&\sum^\infty_{k=0}a_kz^k~~~\text{for }z_s\ge z \ge 0
\eqn
and
\bqn
\lb{HH3}
\frac{A_1(z)}{A_2(z)}&=&\sum^\infty_{k=0}\hat{\beta}_k(z-z_s)^k\nb\\
\frac{A_0(z)}{A_2(z)}&=&\sum^\infty_{k=0}\hat{\gamma}_k(z-z_s)^k\nb\\
R_\text{out}(z)&=&\sum^\infty_{k=0}b_k(z-z_s)^k~~~\text{for }1\ge z \ge z_s
\eqn
where $z_s=1-\frac{r_h}{r_s}$. Finally, the recurrence relation is given by
\bqn
\lb{HH4}
a_1&=&-\frac{\hat{C}_{-1}}{\hat{B}_{-1}}a_0,\nb\\
a_{k+2}&=&-\frac{1}{(k+2)(k+1+\hat{B}_{-1})}\left\{\hat{C}_{-1}a_{k+1}\right.\nb\\
&&\left.+\sum_{i=0}^k\left[(k-i+1)\hat{B}_ia_{k-i+1}+\hat{C}_ia_{k-i}\right]\right\},\nb\\
b_0&=&R_\text{out}(z_s)=R_\text{in}(z_s)=\sum_{i=0}a_iz_s^i,\nb\\
b_1&=&\left.\partial_zR_\text{out}(z)\right|_{r=r_s}=\left.\partial_zR_\text{in}(z)\right|_{r=r_s}\nb\\
&=&\sum_{i=0}a_{i+1}(i+1)z_s^{i},\nb\\
b_{k+2}&=&-\frac{\sum^k_{i=0}\left[(k-i+1)\hat{\beta}_ib_{k-i+1}+\hat{\gamma}_ib_{k-i}\right]}{(k+1)(k+2)}
\eqn
with $k\ge 0$. So once the value of $a_0$ is determined, all $b_i=b_i(\omega)$ can be derived. For the sake of simplicity, we set $a_0=1$, so that the frequency $\omega=w r_h$ can be obtained by solving the equation
\bqn
\lb{HH5}
R_\text{out}(1)=\sum^\infty_{k=0}b_k(1-z_s)^k=0
\eqn
Since the preceding series converges, we only need to extend the series up to $N$ order (where $N\gg 1$) in the actual computation, provided that its error stays within a tolerable range.
\subsection{Finite Difference Method}

The Finite Difference Method operates in the time domain, and it depends on the $r_*$ coordinate, so we compute the tortoise coordinate:
\bqn
\lb{FDM1}
r^\text{in}_*&=&r^\text{out}_*(r_s)+\frac{1}{6r_h}\left[2\sqrt{3}\arctan\frac{2r+r_h}{\sqrt{3}r_h}\right.\nb\\
&&-2\sqrt{3}\arctan\frac{2r_s+r_h}{\sqrt{3}r_h}+2\log\frac{r_h-r}{r_h-r_s}\nb\\
&&\left.+\log\frac{r_h^2+r_hr_s+r_s^2}{r_h^2+r_hr+r^2}\right],\nb\\
r^\text{out}_*&=&\frac{\log(r-c_\text{out}r_h)}{c_\text{out}r_h(3-e_\text{out}-e_\text{out}^2-e_\text{out}^3)}\nb\\
&&+\frac{3+6e_\text{out}+10e_\text{out}^2+6e_\text{out}^3+3e_\text{out}^4}{c_\text{out}r_h(3+2e_\text{out}+e_\text{out}^2)^{3/2}(1+2e_\text{out}+3e_\text{out}^2)}\nb\\
&&\times\text{arccot}\frac{c_\text{out}\sqrt{3+2e_\text{out}+3e_\text{out}^2}r_h}{2r+c_\text{out}(1+e_\text{out})r_h}\nb\\
&&+\frac{e_\text{out}^2\log(r-c_\text{out}e_\text{out}r_h)}{c_\text{out}r_h(e_\text{out}-1)(1+2e_\text{out}+3e_\text{out}^2)}\nb\\
&&-\frac{(1+e_\text{out})^3}{2c_\text{out}r_h(3+2e_\text{out}+e_\text{out}^2)(1+2e_\text{out}+3e_\text{out}^2)}\nb\\
&&\times\log\left[r^2+c_\text{out}(1+e_\text{out})r_hr\right.\nb\\
&&\left.+c_\text{out}^2(1+e_\text{out}+e_\text{out}^2)r_h^2\right]-r_C\nb\\
r_C&=&\frac{3+6e_\text{out}+10e_\text{out}^2+6e_\text{out}^3+3e_\text{out}^4}{2c_\text{out}(3+2e_\text{out}+e_\text{out}^2)^{3/2}r_h}\nb\\
&&\times\frac{\pi}{1+2e_\text{out}+3e_\text{out}^2}
\eqn
so $r_*(r\rightarrow\infty)\rightarrow 0$ and $r_*(r\rightarrow r_h)\rightarrow-\infty$.

By taking $t=t_0+i\Delta t$ and $r_*=r_{*0}+j\Delta r_*$, Eq.(\ref{QNMsA}) become finite difference equation
\bqn
\lb{FDM2}
\Psi^{i+1}_j&=&-\Psi^{i-1}_j+\frac{\Delta t^2}{\Delta r_*^2}\left(\Psi^i_{j-1}+\Psi^i_{j+1}\right)\nb\\
&&+\left(2-2\frac{\Delta t^2}{\Delta r_*^2}-\Delta t^2 V_j\right)\Psi^i_j,
\eqn
Let's choose the initial conditions and boundary condition as
\bqn
\lb{FDM3}
&&\Psi(r_*,t_0)=C_A \exp\left[-C_a(r_*-C_b)^2\right],\nb\\
&&\left.\partial_t\Psi(r_*,t)\right|_{t=t_0}=0,\nb\\
&&\left.\Psi(r_*,t)\right|_{r_*=0}=0.
\eqn

The join condition at the grid point $j_s$ of $r_s$ requires $\Psi_{r\rightarrow r_S^+}=\Psi_{r\rightarrow r_S^-}$ and $\partial_r\Psi_{r\rightarrow r_S^+}=\partial_r\Psi_{r\rightarrow r_S^-}$, so it satisfies
\bqn
\lb{FDM4}
\frac{\Psi^i_{j_b}-\Psi^i_{j_b-1}}{\Delta r_*}=\frac{\Psi^i_{j_b+1}-\Psi^i_{j_b}}{\Delta r_*}
\eqn
or
\bqn
\lb{FDM5}
\Psi^i_{j_b}=\frac{\Psi^i_{j_b-1}+\Psi^i_{j_b+1}}{2}.
\eqn

Finally, pay attention to the Von Neumann stability $\frac{\Delta t^2}{\Delta r_*^2}+\frac{\Delta t^2}{4}V_\text{max}<1$\cite{FDM3,FDM4,FDM5}, so we choose $\Delta r_*=2\Delta t$. By using the method, we can obtain all functions at grid points.

We will employ the preceding three approaches to compute the scalar QNM of the black string with a thin shell in the following section.

\section{Numerical Results}
\renewcommand{\theequation}{4.\arabic{equation}} \setcounter{equation}{0}

We find that MM and GHHM produce highly precise outcomes, so for the sake of simplicity, we assign the step value $1/33$ for MM, so the master differential equations transform into two eigenmatriices with 34 rows and 34 columns. For GHHM, we extend all series functions to 40 order.

\begin{table*}[ht]
\caption{\label{TableI} Scalar Quasinormal modes frequency $\omega$ with $r_p=1$, $r_s=2$, $L=k=0$ and $n=0$}
\centering
\begin{tabular}{c c c c c c}
         \hline\hline
$c_\text{out}$&$e_\text{out}$& $2M$&$2|Q|$ &  \text{MM}&   \text{GHHM}    \\
        \hline
 $1$   &$0$   & $0.5$     &$0$        & $1.84942131358547 - 2.66385266115268i$ & $1.84942131254772 - 2.66385266044390i$\\
 $1$   &$0.1$ & $0.5555$  &$0.333167$ & $1.84886115991618 - 2.66226082053395i$ & $1.84886118067628 - 2.66226081764372i$\\
 $1$   &$0.2$ & $0.624$   &$0.497996$ & $1.84815838246885 - 2.66028120740985i$ & $1.84815842249751 - 2.66028121314482i$\\
 $1$   &$0.3$ & $0.7085$  &$0.645755$ & $1.84727369483746 - 2.65781614263610i$ & $1.84727374309536 - 2.65781617236739i$\\
 $1$   &$0.4$ & $0.812$   &$0.789937$ & $1.84616262788783 - 2.65476139504655i$ & $1.84616265506010 - 2.65476145891629i$\\
 $1.01$&$0$   & $0.515151$&$0$        & $1.84874708634214 - 2.66280679611562i$ & $1.84874708059789 - 2.66280678415070i$\\
 $1.01$&$0.1$ & $0.572332$&$0.339863$ & $1.84818249987103 - 2.66118467764087i$ & $1.84818254560525 - 2.66118465702376i$\\
 $1.01$&$0.2$ & $0.642908$&$0.508006$ & $1.84747391534033 - 2.65916691063215i$ & $1.84747401094012 - 2.65916690900099i$\\
 $1.01$&$0.3$ & $0.729968$&$0.658735$ & $1.84658153010992 - 2.65665355321889i$ & $1.84658165205779 - 2.65665360895264i$\\
 $1.01$&$0.4$ & $0.836604$&$0.805814$ & $1.84546018651184 - 2.65353774273786i$ & $1.84546026737753 - 2.65353788541905i$\\
 $1.05$&$0$   & $0.578813$&$0$        & $1.84586193207136 - 2.65839380268179i$ & $1.84586073437129 - 2.65839231903923i$\\
 $1.05$&$0.1$ & $0.643061$&$0.367316$ & $1.84528415854500 - 2.65665090542783i$ & $1.84528476583011 - 2.65664885481842i$\\
 $1.05$&$0.2$ & $0.722358$&$0.549041$ & $1.84455801780905 - 2.65448060691822i$ & $1.84456061173763 - 2.65447884008743i$\\
 $1.05$&$0.3$ & $0.820177$&$0.711945$ & $1.84364192918572 - 2.65177365694817i$ & $1.84364613137428 - 2.65177352308678i$\\
 $1.05$&$0.4$ & $0.939992$&$0.870905$ & $1.84248828077456 - 2.64841223044930i$ & $1.84249236789830 - 2.64841522044087i$\\
\hline
\hline
\end{tabular}
\end{table*}

\begin{table*}[ht]
\caption{\label{Table} Scalar Quasinormal modes frequency $\omega$ by Matrix Method with $r_p=1$, $r_s=2$, $L=k=0$ and $n=0$}
\centering
\begin{tabular}{c c c c c}
         \hline\hline
 &$e_\text{out}=0$& $e_\text{out}=0.1$ &  $e_\text{out}=0.2$&   $e_\text{out}=0.3$    \\
        \hline
$c_\text{out}=1$~~~~ &$1.84942 - 2.66385i$&$1.84886 - 2.66226i$&$1.84816 - 2.66028i$&$1.84727 - 2.65782i$\\
$c_\text{out}=1.01$ &$1.84875 - 2.66281i$&$1.84818 - 2.66118i$&$1.84747 - 2.65917i$&$1.84658 - 2.65665i$\\
$c_\text{out}=1.03$ &$1.84734 - 2.66065i$&$1.84677 - 2.65896i$&$1.84605 - 2.65687i$&$1.84515 - 2.65426i$\\
$c_\text{out}=1.05$ &$1.84586 - 2.65839i$&$1.84528 - 2.65665i$&$1.84456 - 2.65448i$&$1.84364 - 2.65177i$\\
\hline
\hline
\end{tabular}
\end{table*}

\begin{figure}[tbp]
\centering
\includegraphics[width=0.7\columnwidth]{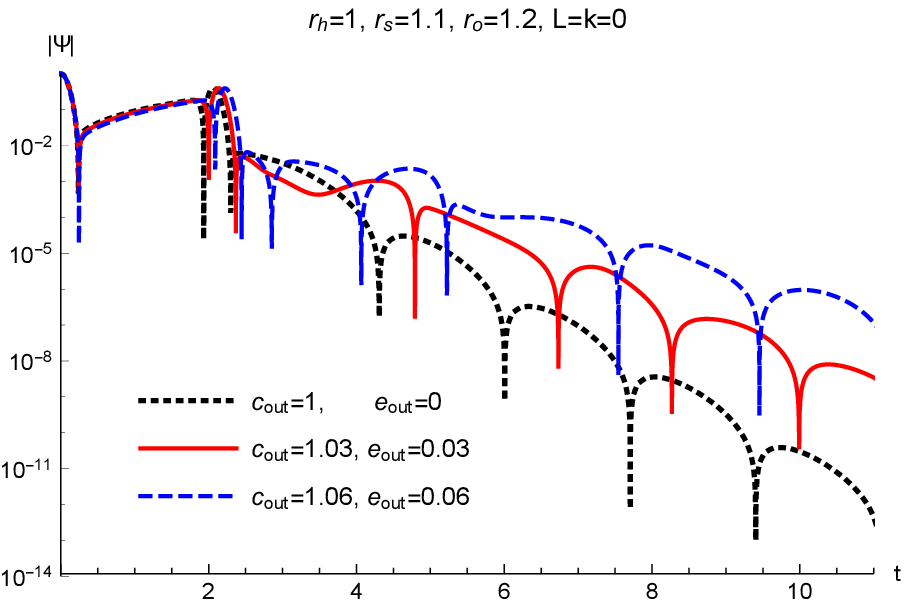}
\includegraphics[width=0.7\columnwidth]{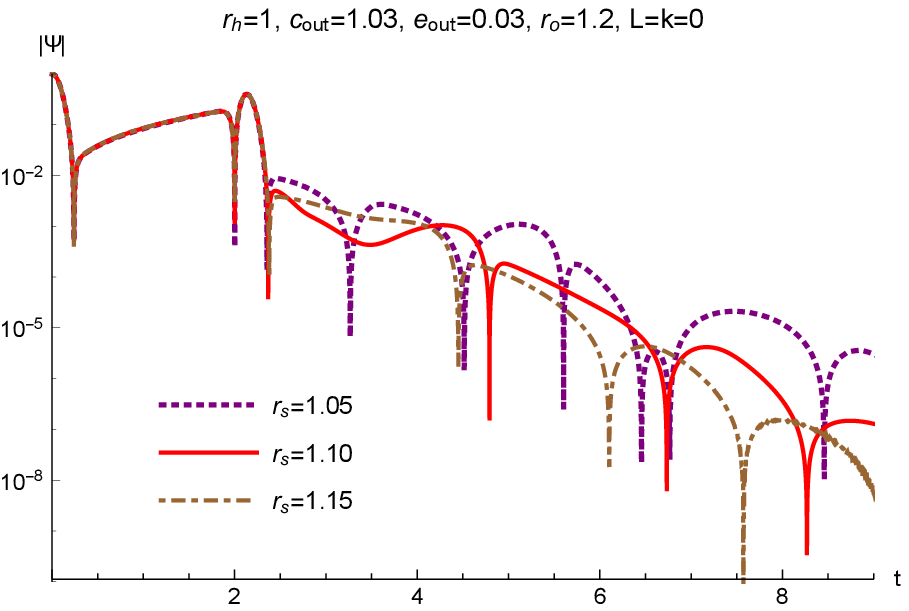}
\caption{Scalar Quasinormal modes with $L=k=0$. The observation point situated at $r_o=1.2$}
\lb{Fig2}
\end{figure}

In Table I, we compare MM with GHHM and observe that they concur with each other at a very high precision, which indicates that both methods are reliable. In Table II, we utilize MM to compute the QNM frequency for various values of $c_\text{out}$ and $e_\text{out}$. We ascertain that when $c_\text{out}$ and $e_\text{out}$ are small, as they increase, both the real part and the absolute value of the imaginary part of QNM frequency will diminish, which implies that QNM oscillation frequency ascends and its amplitude decay rate declines. In FIG.2., we employ FDM to investigate the dynamical process of scalar QNM oscillation, and confirm that they are consistent with our previous results.

We acknowledge that the spacetime outside the thin shell obeys $c_\text{out}<b$. However, once $c_\text{out}=b$ is violated, the thin shell will collapse into a black hole. $e_\text{out}<1$ is determined by charge. $f_\text{in}(r\rightarrow\infty)=f_\text{out}(r\rightarrow\infty)=1$ holds true, but the location of $f_\text{out}=0$ exceeds the location of $f_\text{out}=0$, creating a gap near the discontinuous point $r_s$. As $c_\text{out}/b$ and $e_\text{out}$ approach 1, the gap enlarges significantly, resulting in an echo effect in QNM oscillation\cite{echo1,echo2,echo3}. Interestingly, as these two values approach 1, the beat period of the echo effect lengthens. We illustrate this dynamic process in FIG.3.

\begin{figure*}[tbp]
\centering
\includegraphics[width=0.7\columnwidth]{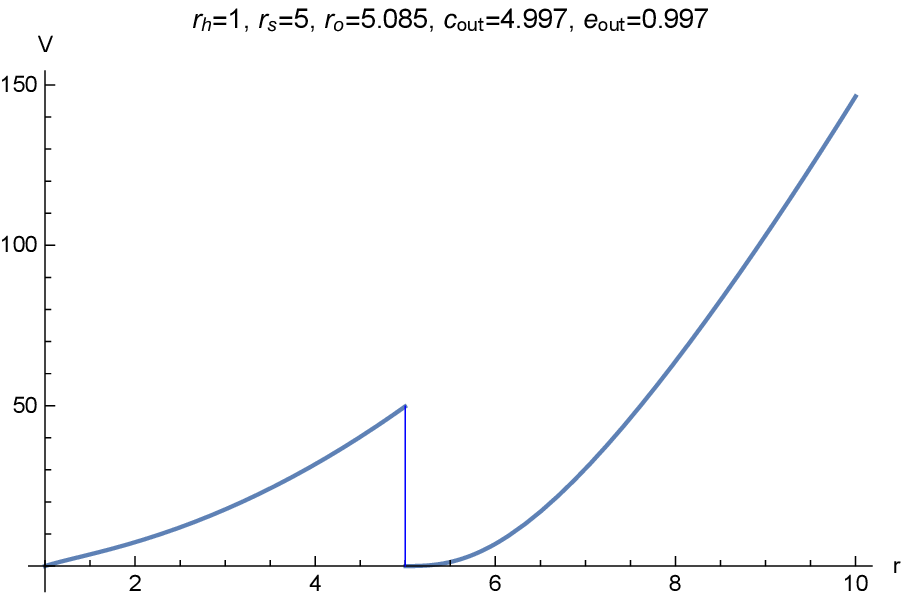}\includegraphics[width=0.7\columnwidth]{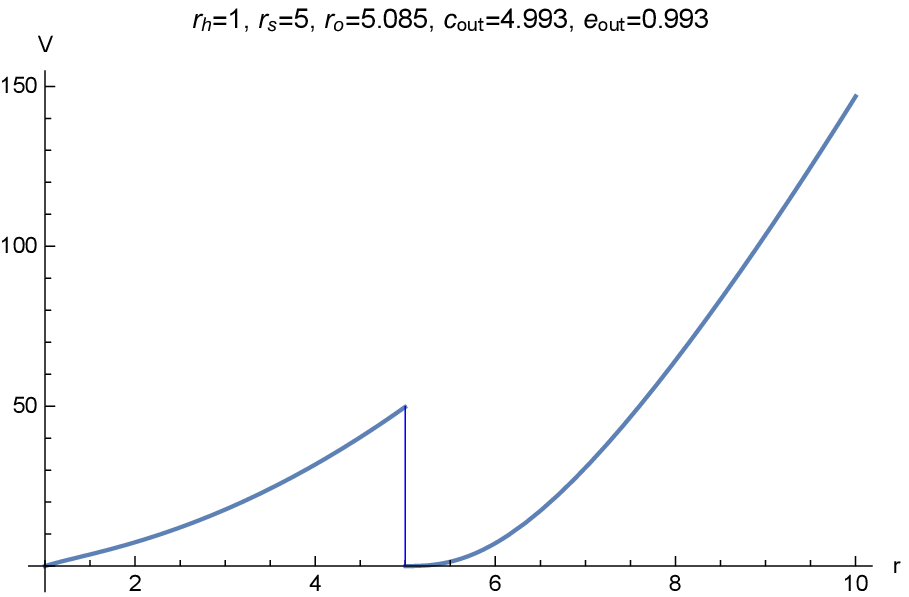}\includegraphics[width=0.7\columnwidth]{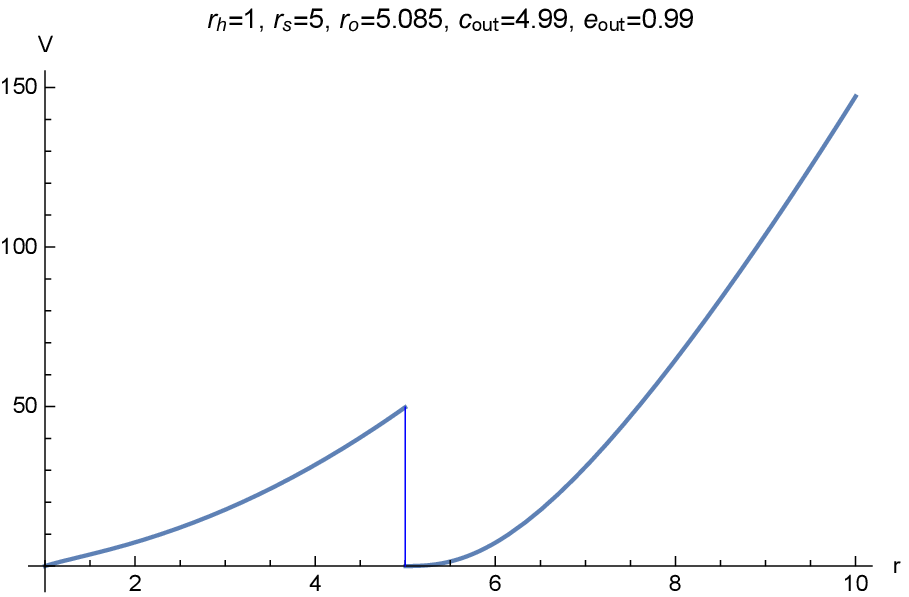}
\includegraphics[width=0.7\columnwidth]{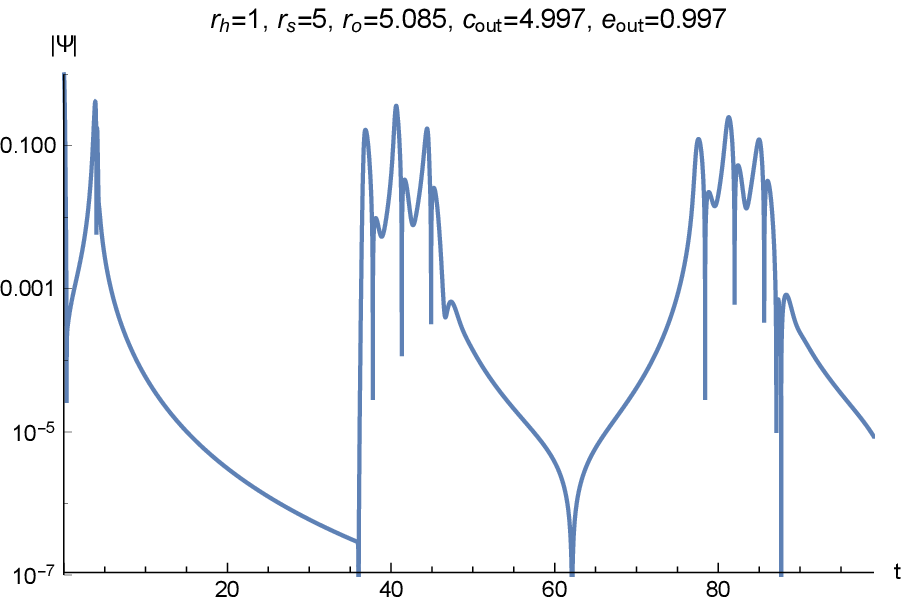}\includegraphics[width=0.7\columnwidth]{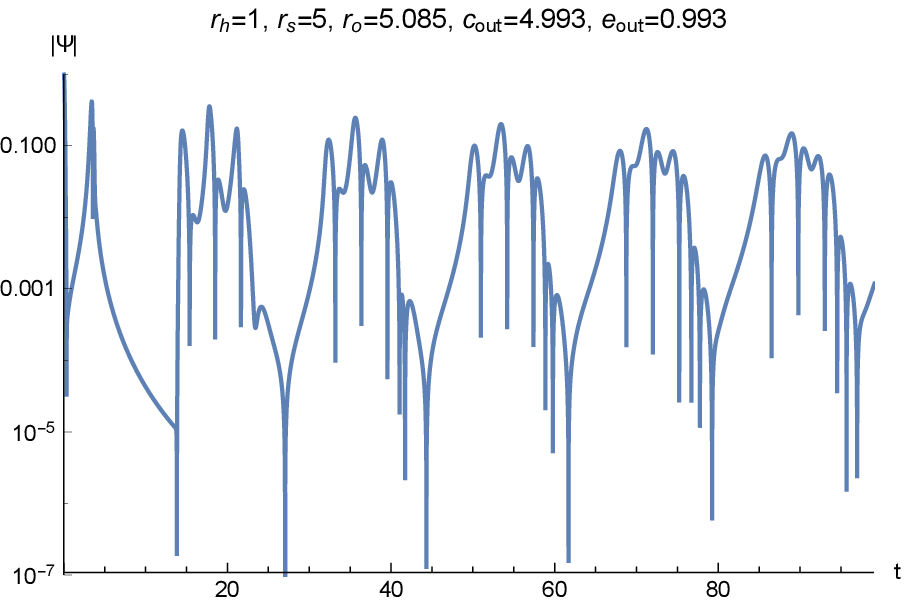}\includegraphics[width=0.7\columnwidth]{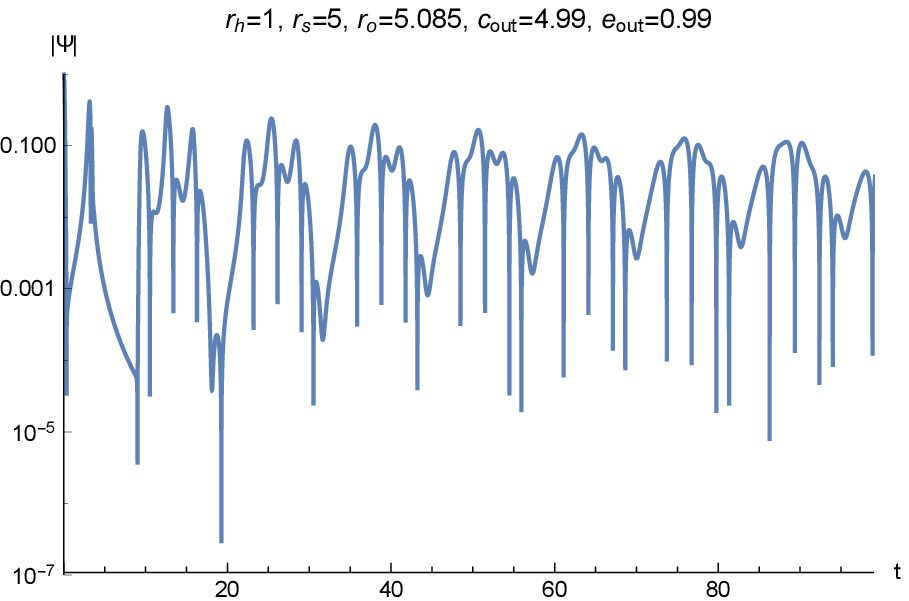}
\caption{Echo effect of the near extreme black hole case with $L=k=0$. The observation point is situated at $r_o=5.085$.}
\lb{Fig3}
\end{figure*}

It is interesting that the curve will change from Echo phase to QNM phase by reducing the mass and charge of the shell, because the gap of potentials become narrower as the parameters decrease. What's more, It is also found that the Echo phase is changed from a \textbf{stable} QNM phase, so the Echo phase decays in amplitude as $t\rightarrow\infty$ (though the Echo duration grows significantly due to the reflection of the wave in the gap). The reason is that the boundary condition requires that the wave function must vanish at infinity, which means that the energy cannot leak at infinity; but from the event horizon, the energy of the gravitational waves will gradually enter the interior of the black string due to the characteristic that classical black holes only absorb radiation and do not emit it, so that the Echo effect outside the black string will slowly decrease with time. The details are shown in FIG4.

\begin{figure*}[tbp]
\centering
\includegraphics[width=0.7\columnwidth]{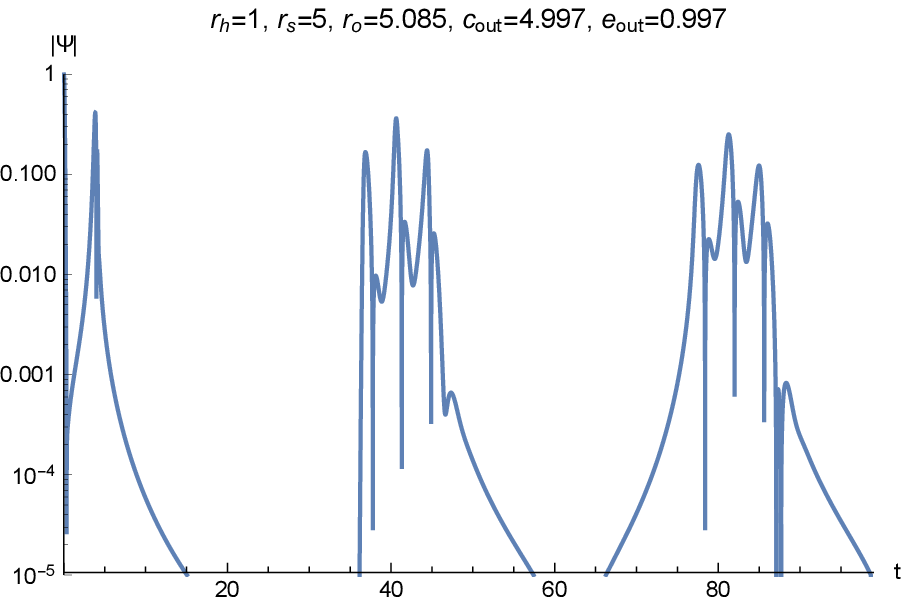}\includegraphics[width=0.7\columnwidth]{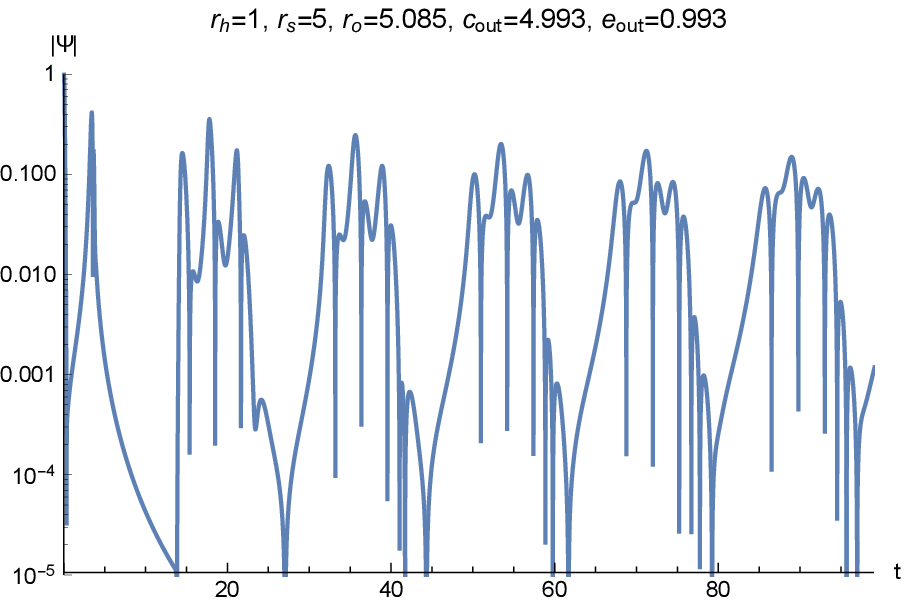}\includegraphics[width=0.7\columnwidth]{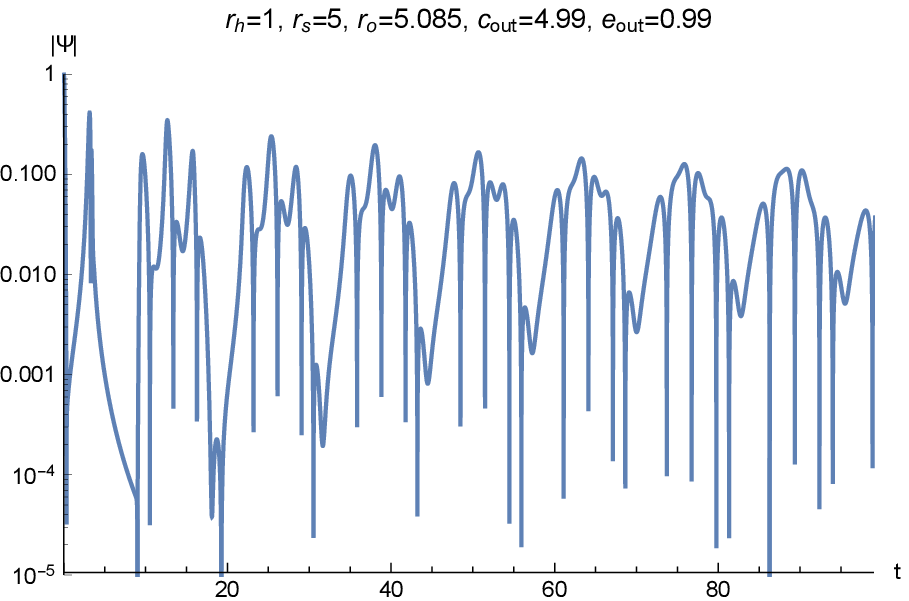}
\includegraphics[width=0.7\columnwidth]{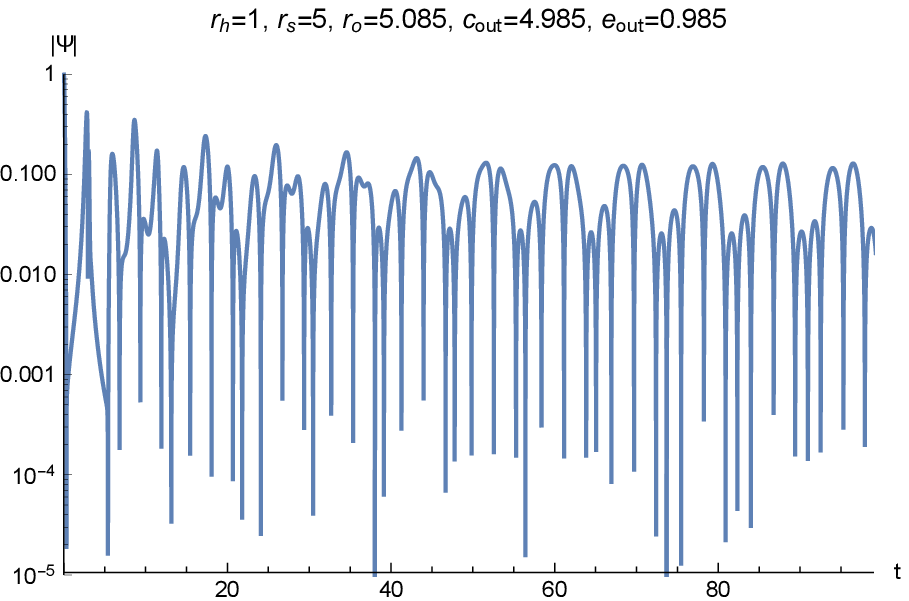}\includegraphics[width=0.7\columnwidth]{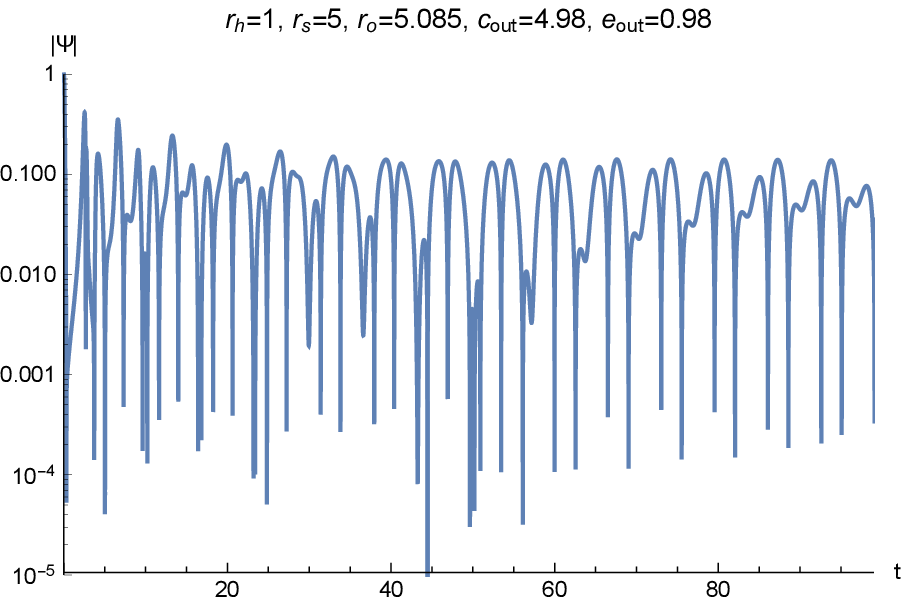}\includegraphics[width=0.7\columnwidth]{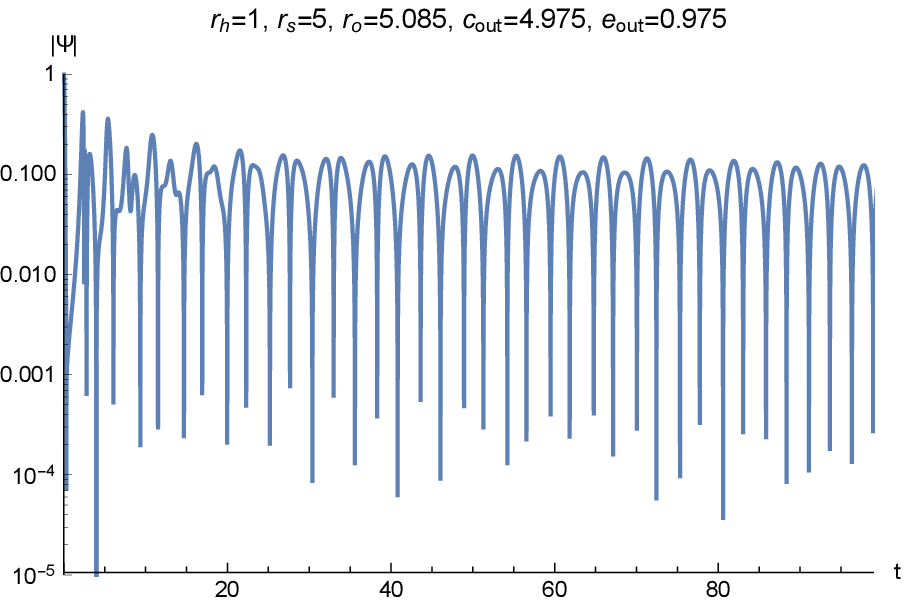}
\includegraphics[width=0.7\columnwidth]{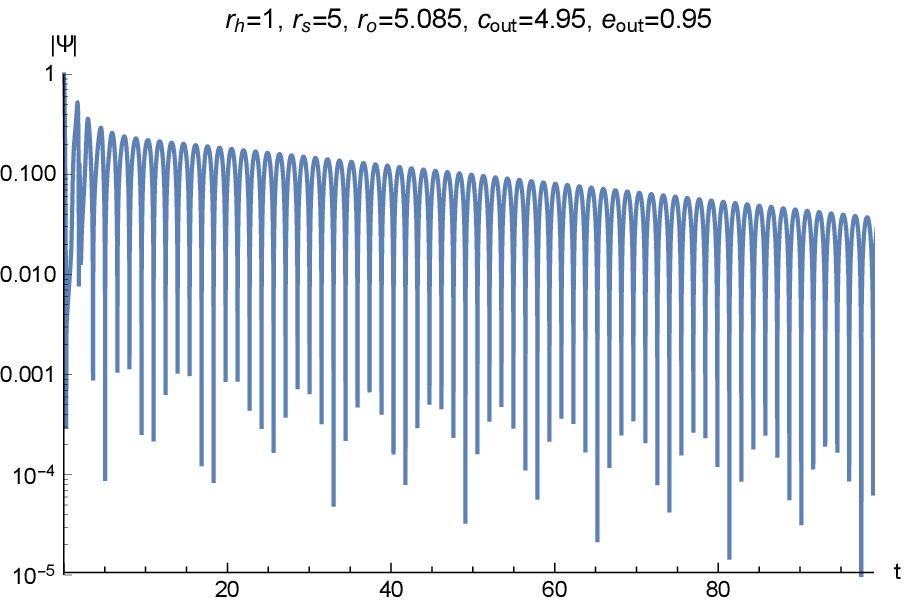}\includegraphics[width=0.7\columnwidth]{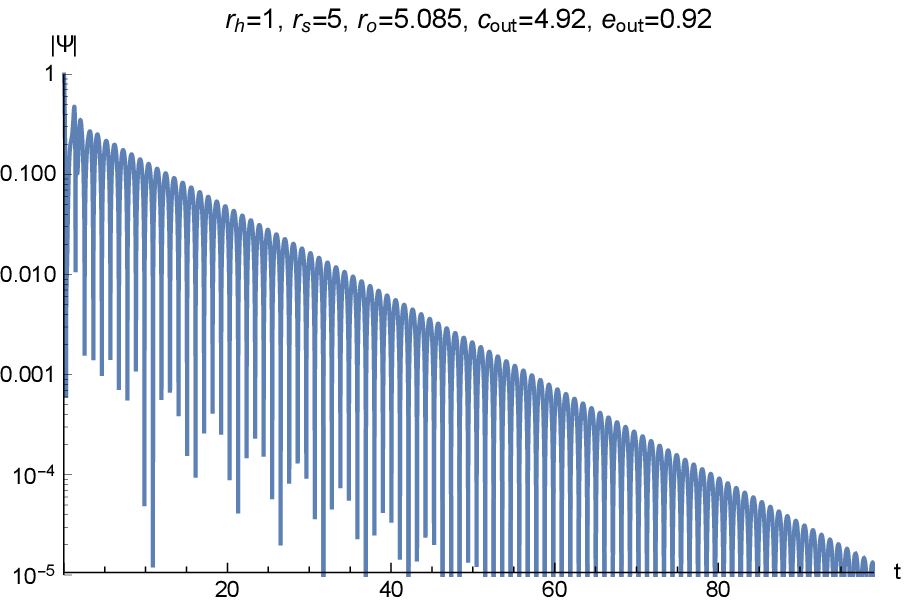}\includegraphics[width=0.7\columnwidth]{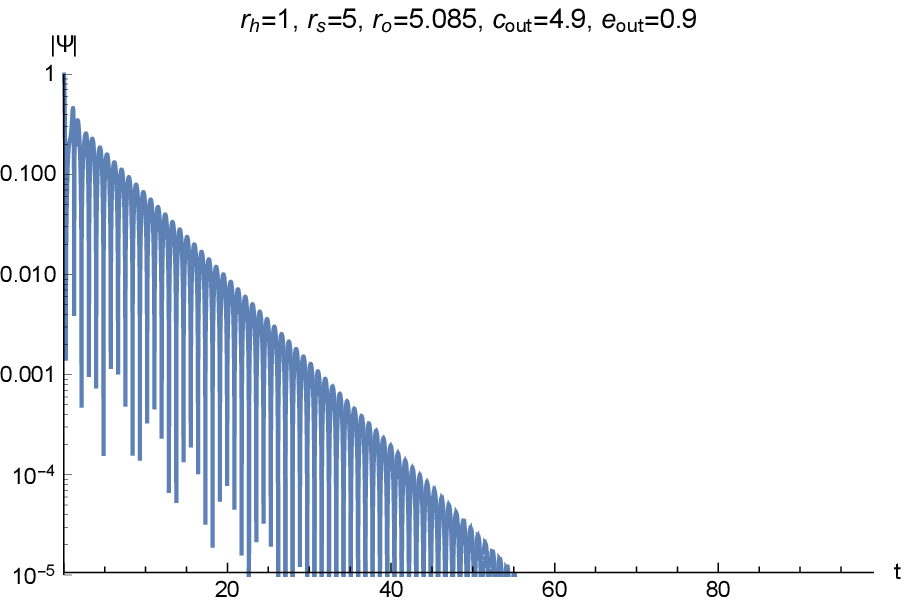}
\caption{From echo phase to QNM phase of the black hole as the shell's mass and charge reduce. The observation point is situated at $r_o=5.085$.}
\lb{Fig4}
\end{figure*}

\section{Conclusion}
\renewcommand{\theequation}{5.\arabic{equation}} \setcounter{equation}{0}

This paper proposes two numerical approaches for computing QNM frequencies of discontinuous potentials: specifically, matrix method and generalized Horowitz-Hubeny method.

Using the matrix method, we map the domain range of both regions inside and outside the thin shell to the interval $[0,1]$ by applying the coordinate transformation (\ref{MM1}) and (\ref{MM3}). Then we approximate the derivative at each grid point as a linear combination of function values at nearby points by using high-accuracy difference, the Runge-Kutta method or the differential quadrature method. This yields two matrix equations. We impose the Wronskian condition as a join condition at the discontinuity point and replace the rows of the matrix corresponding to the thin shell point with this condition after discretization. This results in two eigenmatrices whose eigenvalues and eigenvectors are the QNM frequencies and waveforms. However, similar to the original version of the matrix method, this method faces the problem of excessively large number of matrix rows due to excessive grid points. This leads to significantly longer computation time when the number of grid points exceeds 50. To overcome this problem, we propose a technique that combines high-order matrix method with secant method for computing QNM\cite{Lin4}. In the work involving discontinuous potential functions, we have to face the problem of solving multiple determinants simultaneously. However, this can be calculated efficiently by using high-order matrix method and the Broyden method.

Using the generalized Horowitz-Hubeny method, we simply require that the function and its first-order derivatives of the two master equations are equal at the discontinuity point in AdS black hole spacetime. In fact, we can apply the same logic to solve QNM of asymptotically flat spacetime and de Sitter spacetime: we just need to set $R(r)=e^{2i\omega r}r^{2i\omega r_h}\Psi(r)$ and $x=1-\frac{r_h}{r}$ in Eq.(\ref{QNMsC}) for the asymptotically flat spacetime case, or $R(r)=(r_c-r)^{2i\omega/f'(r_c)}\Psi(r)$ and $x=\frac{r-r_h}{r_c-r_h}$ for the de Sitter spacetime with cosmological horizon $r_c$, so that the master equation becomes
\bqn
\lb{HHF1}
A_2(x)\Psi''(x)+A_1(x)\Psi'(x)+A_0(x)\Psi(x)=0.
\eqn
Then, we can expand $\frac{A_1}{A_2}=\sum_i \hat{a}_i x^i$, $\frac{A_0}{A_2}=\sum_i \hat{b}_i x^i$ and $\Psi(x)=\sum_i \hat{c}_i x^i$ near the horizon $x=0$, and deduce the recurrence relation of $\hat{a}_i$, $\hat{b}_i$ and $\hat{c}_i$. Subsequently, the QNM frequency is derived from the boundary condition $\Psi'(x=1)\rightarrow0$. Unlike CFM, GHHM does not necessitate that $A_2(x)$, $A_1(x)$, $A_0(x)$ are rational expressions, thus GHHM has a broader scope of application.

We apply the aforementioned approach to investigate scalar quasinormal modes of a rotating cylindrical AdS black hole with a thin shell, and also discover that the discontinuous potential function $V(r)$ exhibits a large gap in the case of a very massive and strongly charged thin shell, resulting in an echo effect in the quasinormal mode wave. Using the finite difference method, we obtain the echo wave and observe a significant difference from the echo waveform in asymptotically flat black hole spacetime.

In fact, a black hole with a thin shell is the simplest example of a black hole with a discontinuous potential function. More realistic cases could involve black holes with an accretion disk or dark matter, and the research on the ringdown gravitational wave of such realistic black hole spacetimes has more practical significance. We will pursue this open problem in the future.

\begin{acknowledgments}
This work was supported by
National Natural Science Foundation of China (NNSFC) under contract No. 42230207.
The author would like to thank Jian Zhang and Meifang Chen for improving the English.
The author would like to express his gratitude to Hong-sheng Zhang for helpful discussion.
\end{acknowledgments}


\begin{thebibliography}{nbound}

\bibitem{LIGO1} B. P. Abbott et al. (LIGO Scientific and Virgo Collaborations), Phys. Rev. Lett. 116, 061102 (2016).
\bibitem{LIGO2} B. P. Abbott et al. (LIGO Scientific and Virgo Collaborations), Phys. Rev. Lett. 116, 241103 (2016).
\bibitem{LIGO3} B. P. Abbott et al. (LIGO Scientific and Virgo Collaborations), Phys. Rev. Lett. 118, 221101 (2017).
\bibitem{LIGO4} B. P. Abbott et al. (LIGO Scientific and Virgo Collaborations), Astrophys. J. 851, L35 (2017).
\bibitem{LIGO5} B. P. Abbott et al. (LIGO Scientific and Virgo Collaborations), Phys. Rev. Lett. 119, 141101 (2017).
\bibitem{LIGO6} B. P. Abbott et al. (LIGO Scientific and Virgo Collaborations), Phys. Rev. Lett. 119, 161101 (2017).
\bibitem{LIGO7} B. P. Abbott et al. (LIGO Scientific and Virgo Collaborations), Phys. Rev. X 6, 041015 (2016).
\bibitem{LIGO8} F. Acernese et al., Classical Quantum Gravity 32, 024001 (2015).
\bibitem{LIGO9} B. P. Abbott et al. (Virgo, Fermi-GBM, INTEGRAL, and LIGO Scientific Collaborations), Astrophys. J. Lett. 848, L13 (2017).


\bibitem{WKB1} B. F. Schutz and C. M. Will, Astrophysical Journal, 291, L33-L36 (1985)
\bibitem{WKB2} S. Iyer and C. M. Will, Phys. Rev. D35, 3621 (1987)
\bibitem{WKB3} R. A. Konoplya, Phys. Rev. D68, 024018 (2003) arXiv:gr-qc/0303052]
\bibitem{WKB4} S. H. V\"{o}lkel and K.D. Kokkotas, Class.Quant.Grav. 34, 125006(2017).

\bibitem{CFM} E. Leaver, Proceedings of the Royal Society A402, 285-298(1985)

\bibitem{AIM1} H. T. Cho et al., Class. Quantum Gravity 27, 155004 (2010). arXiv:0912.2740
\bibitem{AIM2} H. T. Cho et al., Adv. Math. Phys., 281705 (2012). arXiv:1111.5024
\bibitem{AIM3} H. T. Cho, J. Doukas, W. Naylor and A.S. Cornell, Phys. Rev. D 83, 124034 (2011). arXiv:1104.1281

\bibitem{HHM} G. T. Horowitz and V. E. Hubeny, Phys. Rev. D62, 024027 (2000)

\bibitem{Lin1} K. Lin and W.-L. Qian, 2016 A non-grid-based interpolation scheme for the eigenvalue problem (arXiv:1609.05948)
\bibitem{Lin2} K. Lin and W.-L. Qian, Class. Quantum Grav. 34, 095004 (2017) arXiv:1610.08135
\bibitem{Lin3} K. Lin and W.-L. Qian, Mod. Phys. Lett. A32, 1750134 (2017) arXiv:1703.06439
\bibitem{Lin4} K. Lin and W.-L. Qian, Chin. Phys. C43,035105 (2019) arXiv:1902.08352
\bibitem{Lin5} K. Lin, Y. Liu, W.L. Qian, B. Wang, and E. Abdalla, Phys. Rev. D100, 065018 (2019) arXiv:1909.04347
\bibitem{Lin6} K. Lin, Y.-Y. Sun and H. Zhang, Phys. Rev. D103, 084015 (2021) arXiv:2104.06631
\bibitem{Lin7} K. Lin and W.-L. Qian, 2022 High-order matrix method with delimited expansion domain
(arXiv:2209.11612)


\bibitem{FDM1} C. Gundlach, R. H. Price, J. Pullin, Phys.Rev. D49, 883-889 (1994) 	arXiv:gr-qc/9307009
\bibitem{FDM2} C. Gundlach, R. H. Price, J. Pullin, Phys.Rev. D49, 890-899 (1994) arXiv:gr-qc/9307010
\bibitem{FDM3} B. Cuadros-Melgar, J. de Oliveira, and C. E. Pellicer, Phys. Rev. D85, 024014 (2012)
\bibitem{FDM4} E. Abdalla, OwenPavelFernandez Piedra, F.S. Nunez and
J. deOliveira, Phys. Rev. D88, 064035 (2013)
\bibitem{FDM5} B. Cuadros-Melgar, J. de Oliveira, and C. E. Pellicer, J. Phys. Conf. Ser. 453, 012025 (2013)



\bibitem{metric} J. P. S. Lemos and V. T. Zanchin, Phys. Rev. D54, 3840 (1996) arXiv:hep-th/9511188



\bibitem{echo1} V. Cardoso, E. Franzin, and P. Pani, Phys. Rev. Lett. 116, 171101 (2016), [Erratum:
Phys.Rev.Lett. 117, 089902(E) (2016)], arXiv:1602.07309 [gr-qc]
\bibitem{echo2} K. Lin and W.-L. Qian, 2022 Echoes in star quasinormal modes using an alternative finite difference method (arXiv:2204.09531)
\bibitem{echo3} K. Lin et al., 2023 Echoes of Stars in tracking Rastall gravity (arXiv:2302.10875)








\end{thebibliography}
\end{document}